\documentclass{article}

\usepackage[dblblindworkshop, final]{neurips_2026}

\usepackage[utf8]{inputenc}
\usepackage[T1]{fontenc}
\usepackage{hyperref}
\usepackage{url}
\usepackage{booktabs}
\usepackage{amsfonts}
\usepackage{amsmath}
\usepackage{nicefrac}
\usepackage{microtype}
\usepackage{xcolor}
\usepackage{graphicx}
\usepackage{algorithm}
\usepackage{algpseudocode}
\usepackage{siunitx}
\usepackage{algorithm}
\usepackage{algpseudocode}

\title{Retrieve-then-Disambiguate: Scalable Reviewer Discovery}

\author{
  Vicente Amado Olivo \\
  Michigan State University \\
  \And
  Joe Petrecca \\
  Michigan State University \\
  \And
  Wolfgang E. Kerzendorf \\
  Michigan State University
}

\begin{document}
\maketitle

\begin{abstract}
The growing number of scientific submissions is straining the peer review process, while a large pool of relevant researchers remains unidentified and is therefore not considered as potential reviewers. We formulate the discovery of these researchers as a retrieval problem: given a submission, we retrieve and rank researchers based on domain similarity. We present retrieve-then-disambiguate, an open-source algorithm that identifies domain relevant researchers for a query submission by retrieving similar publications from the scientific literature and resolving ambiguous author names into unique researcher identities. We evaluate the algorithm on \num{500} astronomy and physics publications. We use ORCID-linked first authors, coauthors, and cited researchers as proxy labels to evaluate the recovery of researchers with known connections to each query publication. When retrieving the top \num{500} most similar papers, the algorithm recovers roughly 68\% of first authors, 60\% of coauthors, and 45\% of cited researchers. Our code is available at \url{https://github.com/deepthought-initiative/Retrieve-then-Disambiguate}.
\end{abstract}

\section{Introduction}\label{sec:intro}
The exponential rise in submissions to conferences, journals, and funding agencies has placed growing pressure on the peer review system \citep{merrifield_telescope_2009, leyton-brown_matching_2022, Hanson2023TheSO}. Advances in large language models have accelerated this rise, for example, with NeurIPS submissions growing from \num{9467} in 2020 to \num{21575} in 2025 \citep{neurips2025reflections}. In response, scientific organizations are deploying automated systems across the peer review process, for example, to assign peer reviewers \citep{charlin_framework_2012, kerzendorf_distributed_2020, stelmakh_gold_2025, noauthor_openreviewopenreview-expertise_2025}. In practice, many fields across science still assign peer reviewers manually \citep{merrifield_telescope_2009,jerabkova_first_2023}. \citet{Kovanis2016TheGB} show this manual process produces a distribution problem, not a supply problem, with a minority of researchers reviewing the majority of submissions. This highlights the need for scalable methods to identify domain relevant researchers beyond those already known to editors and program committees \citep{merrifield_telescope_2009, stelmakh2021novice,doi:10.1073/pnas.2507394122}.

Identifying and assigning peer reviewers is a multiobjective problem, involving domain similarity, conflicts of interest, career-stage, and other considerations \citep{langfeldt2001decision, nightingale2007peer, latypova2023decision}. In practice, reviewer assignment systems at large conferences or scientific organizations typically involve two stages: estimating domain relevance between submissions and potential reviewers using vector representations \citep{stelmakh_gold_2025, olivo2026traditionalstatisticalrepresentationsoutperform} and assigning reviewers using optimization algorithms based on a set of constraints \citep{stelmakh_peerreview4all_2019}. The relative importance of these objectives and constraints depends on the priorities of editors and program committees. We focus on the specific challenge of identifying researchers with domain/topical similarity to a submission \citep{mimno_expertise_2007, jerabkova_first_2023, stelmakh_gold_2025}. Currently deployed reviewer assignment systems match or rank reviewers within a predefined pool of candidates \citep[e.g., the Toronto Paper Matching System or OpenReview][]{charlin_framework_2012, noauthor_openreviewopenreview-expertise_2025}. Furthermore, some journals or conferences rely on proprietary reviewer recommendation systems \footnote{\url{https://www.elsevier.com/editor/submission-systems/find-reviewers-tips}}, whose data or matching methodology is not openly published.

Therefore, we present an open source algorithm for identifying relevant reviewers directly from the scientific literature. We formulate reviewer discovery as an information retrieval problem: given a query submission, retrieve and rank researchers by domain similarity. We present the retrieve-then-disambiguate algorithm: (1) retrieve domain similar papers to the submission, (2) disambiguate author identities within the retrieved pool, and (3) rank the disambiguated identities by similarity to the submission. We leverage authorship, coauthorship, and citation relationships as observable signals of domain similarity. Although these relationships do not establish reviewer suitability, they allow us to evaluate whether a retrieval system successfully identifies researchers within the domain of the query publication. In summary, we introduce an open-source algorithm for identifying topically relevant researchers directly from the scientific literature, without relying on a predefined candidate pool. Second, we present an evaluation using astronomy and physics publications from NASA ADS, with authorship, coauthorship, and citation relationships as signals of domain relevance.

\algrenewcommand\algorithmicrequire{\textbf{Input:}}
\algrenewcommand\algorithmicensure{\textbf{Output:}}
\begin{algorithm}[t!]
\caption{Retrieve-then-Disambiguate}
\label{alg:pipeline}
\begin{algorithmic}
\Require Query paper $q$, corpus $\mathcal{D}$, pool size $N$, embedding function $\phi(\cdot)$, disambiguation function $\delta(\cdot)$
\Ensure Ranked list of disambiguated identities

\State \textbf{Stage 1: Retrieval}
\State $\mathcal{D}' \gets \{d \in \mathcal{D} : \operatorname{year}(d) < \operatorname{year}(q)\}$
\State $\mathbf{v}_q \gets \phi(q)$
\For{each $d \in \mathcal{D}'$}
    \State $\mathbf{v}_d \gets \phi(d)$
    \State $s_d \gets \dfrac{\mathbf{v}_q \cdot \mathbf{v}_d}{\|\mathbf{v}_q\| \, \|\mathbf{v}_d\|}$
\EndFor
\State $\mathcal{C} \gets$ top-$N$ papers in $\mathcal{D}'$ by $s_d$

\State \textbf{Stage 2: Name-blocked disambiguation}
\State $M \gets$ all author records across papers in $\mathcal{C}$
\State Partition $M$ into blocks $b_1, \ldots, b_k$ by surname + first initial
\For{$j = 1, \ldots, k$}
    \State $U_j \gets \delta(b_j)$ \Comment{disambiguated identities from block $j$}
\EndFor
\State $\mathcal{I} \gets \bigcup_j U_j$

\State \textbf{Stage 3: Expertise ranking}
\For{each identity $u \in \mathcal{I}$}
    \State $s_u \gets \dfrac{\mathbf{v}_q \cdot \phi(u)}{\|\mathbf{v}_q\| \, \|\phi(u)\|}$ \Comment{$\phi(u)$ over $u$'s concatenated papers}
\EndFor
\State \Return $\mathcal{I}$ sorted by decreasing $s_u$

\end{algorithmic}
\end{algorithm}

\section{Data}
\label{sec:data}
We evaluate the retrieve-then-disambiguate algorithm using publication metadata from astronomy and physics literature linked to ORCID identifiers. We cross-match all refereed article titles in the NASA Astrophysics Data System (ADS)\footnote{\url{https://ui.adsabs.harvard.edu/}} against the 2025 ORCID public data dump\footnote{\url{https://info.orcid.org/documentation/integration-guide/working-with-bulk-data/}}. This yields \num{2282214} unique publications associated with \num{689003} unique ORCID-linked researchers. We randomly sample \num{500} query publications whose first authors have at least \num{20} ORCID-linked publications in the corpus. This restriction reduces cases where retrieval failures may reflect insufficient publication data rather than limitations of the algorithm. For each query publication, we retrieve candidate domain reviewers from a larger corpus of \num{4942260} astronomy and physics publications published between \num{1995} and \num{2024}.

We use the ORCID-linked data to construct three proxy labels for evaluating retrieval: first author (the query paper's first author), coauthor (any other author of the query paper), and citation (any author of a paper cited by the query). These labels provide observable links between domain researchers and query publications but do not fully capture domain expertise \citep{stelmakh_gold_2025, olivo2026traditionalstatisticalrepresentationsoutperform}. We therefore use these labels to evaluate reviewer discovery rather than reviewer expertise.

\section{Methods}
\label{sec:method}
Algorithm~\ref{alg:pipeline} presents the retrieve-then-disambiguate algorithm and we describe each stage below.

\textbf{Stage 1: Retrieval.} Given a query paper's title and abstract, we retrieve the $N$ most similar candidate papers. To mimic the constraints of real peer review, in which only prior work is available, we restrict candidates to papers published before the query paper. We use TF--IDF and cosine similarity to estimate similarity between candidate papers and a query paper. A single TF--IDF vectorizer is fit once over all query paper titles and abstracts \citep{olivo2026traditionalstatisticalrepresentationsoutperform}, using the scikit-learn implementation \citep{pedregosa_scikit-learn_2018} (reused in Stage 3). We set the n-gram range to capture both unigrams and bigrams, allowing the model to incorporate two word phrases (e.g., ``stellar evolution,'' ``galaxy formation,'' or ``white dwarf'') that represent more domain-specific meaning than separate individual words (e.g., ``white,'' ``dwarf''). Standard English stop words (e.g., articles, prepositions) are removed to reduce noise. We compute cosine similarity between the TF--IDF vector representations of the query and each candidate paper, and keep the top $N$ candidates by similarity as the candidate pool for Stage 2.

\textbf{Stage 2: Name-blocked Disambiguation.} Every one of the $N$ retrieved candidate papers' full author lists is separated into individual author records, one row per author with metadata (i.e., author name string, title, abstract, and publication year). Author records are split into disjoint name blocks by first name initial and last name (e.g., `J.Smith' or `Y.Wang'), following standard blocking practice in author name disambiguation \citep{subramanian_s2and_2021}. Within each block, we disambiguate author records following the approach of \citet{olivo2025practicalauthordisambiguationmetadata}. Each author record is embedded by concatenating a character-level author name embedding with a SPECTER2 title/abstract embedding \citep{cohan_specter_2020}, then passing those concatenated embeddings through a trained contrastive encoder to produce a final author record embedding. The embedding representations of each author record within a block are then clustered with a constrained variant of DBSCAN \citep{ester_density-based_1996, khabsa_online_2015}. Each resulting cluster is treated as one disambiguated author identity. For evaluation, we assign ORCID labels to clusters by matching publication titles. An ORCID label is considered retrieved if it appears in a cluster with no conflicting ORCID labels.

\textbf{Stage 3: Expertise ranking.} We construct a disambiguated author representation by concatenating the title and abstract of every distinct paper attributed to a disambiguated cluster. Using the TF--IDF vectorizer, we transform every disambiguated author identity's representation into this shared vector space, and rank disambiguated identities by cosine similarity to the query vector representation.

\begin{table}[t!]
\centering
\caption{ORCID-linked researcher recovery rate by retrieval depth \(N\), evaluated over \num{500} query publications with \num{500} first-author, \num{318} coauthor, and \num{7416} citation labels. Recovery is measured across all disambiguated candidate identities, without applying a fixed-size ranking cutoff. Mean blocks reports the average number of name blocks per query.}
\label{tab:hitrate}
\begin{tabular}{@{}rcccc@{}}
\toprule
$N$ & First Author & Coauthors & Citation & Mean blocks \\
\midrule
500    & 0.680 & 0.597 & 0.447 & 1{,}789 \\
1{,}000  & 0.722 & 0.623 & 0.538 & 3{,}287 \\
10{,}000 & 0.844 & 0.711 & 0.789 & 21{,}922 \\
\bottomrule
\end{tabular}
\end{table}

\section{Results}
\label{sec:results}
\begin{figure*}[t!]
    \centering
    \includegraphics[width=\textwidth]{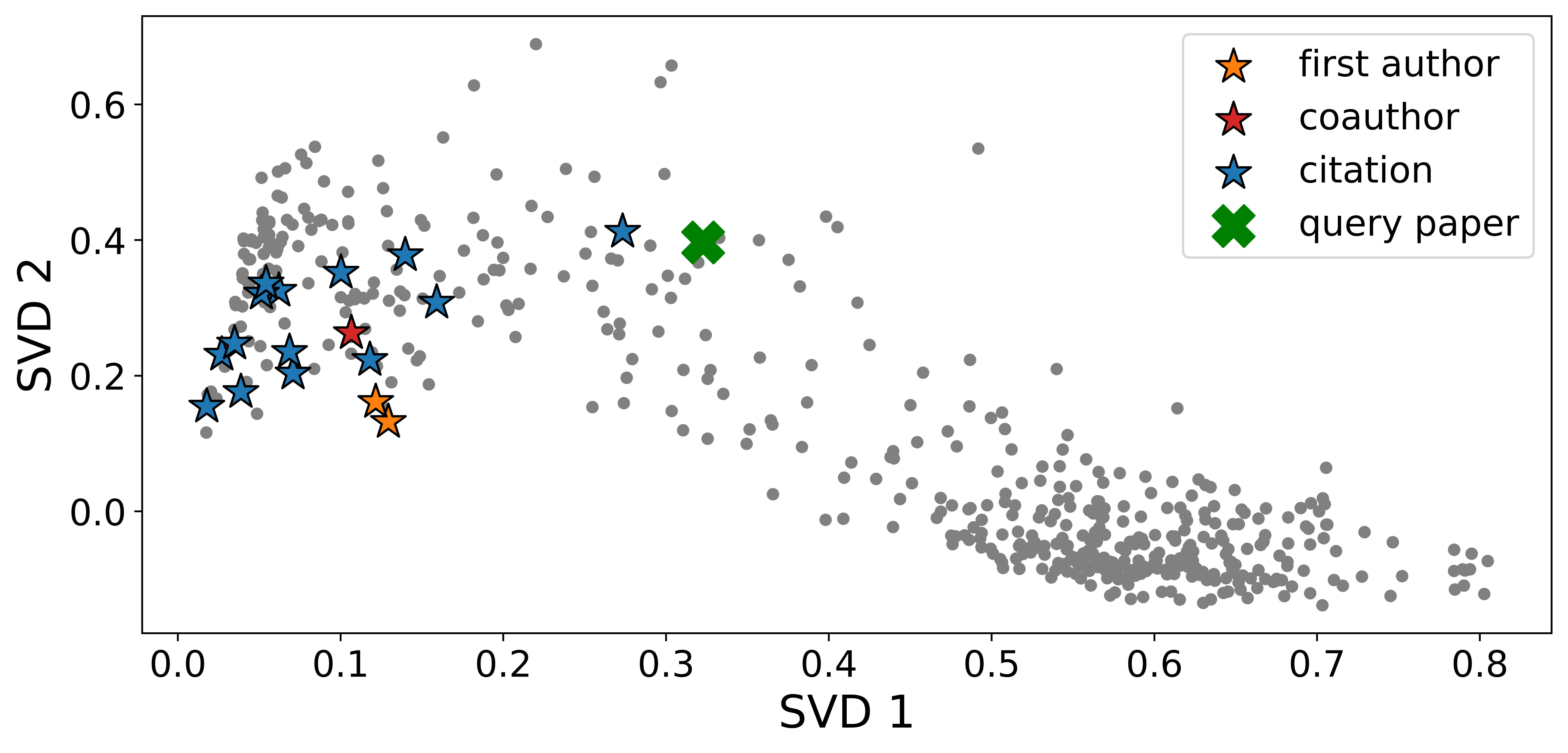}
    \caption{Truncated SVD projection of TF--IDF vector representations for the query paper: "Induced Polarization as a Tool to Assess Mineral Deposits: A Review".}
    \label{fig.sim_space}
\end{figure*}

We evaluate reviewer discovery by measuring the retrieval of ORCID-linked researchers with observable connections to a query publication. A researcher is counted as retrieved if their ORCID is among the disambiguated candidates retrieved. Because the number of candidate identities increases with \(N\), our evaluation measures candidate discovery rather than the quality of a fixed-size reviewer shortlist. Table~\ref{tab:hitrate} reports the hit rate for each label type as we vary the top $N$ publications retrieved. The citation hit rate increases with $N$ from 0.447 when retrieving \num{500} publications to 0.789 when retrieving \num{10000} publications, while first author and coauthor hit rates increase from 0.680 to 0.844 and from 0.597 to 0.711, respectively. Meanwhile, mean name blocks per query grow with $N$, from \num{1789} at $N{=}500$ to \num{21922} at $N{=}10000$, directly increasing disambiguation cost. Figure~\ref{fig.sim_space} visualizes the resulting vector space for one query paper, using $N=500$.

We diagnose failures to recover ORCID-linked researchers across the \num{500} query publications. For each missed labeled researcher, we determine whether they have publications in the corpus preceding the query year. We classify failures as cold start (no prior publications), retrieval miss (prior publications exist but none appear among the top \(N\) retrieved papers), or disambiguation noise (at least one prior publication was retrieved, but identity was not resolved). Across all three retrieval settings, we identify \num{4337} labeled retrieval failures. Of these, 98.2\% were retrieval misses and 1.8\% were cold starts, with no failures attributed to disambiguation noise. We profile the algorithm on \num{10} query publications at different retrieval depth using approximately \num{15} CPU cores, one GPU with 2.1 GiB peak memory usage, and 54.1 GiB of system RAM. Mean end-to-end runtime, excluding one-time model initialization or embedding costs, increases from 15.95 seconds per query at \(N=500\) to 212.44 seconds at \(N=10000\). At \(N=500\), embedding takes 5.45 seconds per query, followed by retrieval (3.52 seconds) and clustering (2.47 seconds). At \(N=10000\), embedding takes 121.86 seconds per query and clustering takes 41.06 seconds. Runtime increases with the number of retrieved publications and name blocks requiring disambiguation.

\section{Discussion and Future Work}
\label{sec:discussion}
Our results show that retrieving disambiguated authors directly from the literature surfaces domain relevant reviewers for peer review. Citation-based recall improves as the candidate pool $N$ grows, indicating that a share of domain relevant reviewers sits outside what a small, fixed candidate pool can capture \citep{Kovanis2016TheGB}. 

While we evaluate the algorithm on astronomy and physics literature, the architecture is not domain-specific: TF--IDF retrieval can be swapped for alternative representation methods, such as SPECTER2 \citep{cohan_specter_2020} embeddings or more recent large language model embeddings (e.g., Qwen embeddings) \footnote{\url{https://huggingface.co/Qwen/Qwen3-Embedding-4B}}, and the disambiguation stage can be replaced with alternative methods \citep{subramanian_s2and_2021}. Applying the same pipeline to other domains and environments, such as computer science conference submissions, would only require refitting $\phi$ and $\delta$ to that literature, not redesigning the retrieval-then-disambiguate algorithm. 

Our evaluation relies on authorship, coauthorship, and citation relationships as proxy signals for domain relevance. These relationships do not establish reviewer expertise, availability, or the absence of conflicts of interest. Furthermore, our reported hit rates measure the recovery of labeled researchers within the retrieved candidate pool, rather than the quality of a fixed-size ranked reviewer shortlist. Our evaluation is also conducted entirely on a preliminary sample of \num{500} astronomy and physics query papers. Therefore, our future work includes evaluating the pipeline across a larger set of the available ORCID-linked data and extending evaluation to new domains and publishing environments, particularly computer science and machine learning conference publishing (e.g., using OpenReview's public submission and review data). Retrieving reviewers directly from the literature, rather than from a predefined pool, offers a path toward surfacing the qualified researchers who remain unknown under current reviewer assignment practices and systems. 

\bibliographystyle{plainnat}
\bibliography{expertise}

\begin{thebibliography}{24}
\providecommand{\natexlab}[1]{#1}
\providecommand{\url}[1]{\texttt{#1}}
\expandafter\ifx\csname urlstyle\endcsname\relax
  \providecommand{\doi}[1]{doi: #1}\else
  \providecommand{\doi}{doi: \begingroup \urlstyle{rm}\Url}\fi

\bibitem[neu(2025)]{neurips2025reflections}
Reflections on the 2025 review process from the program committee chairs.
\newblock NeurIPS Blog, 2025.
\newblock URL \url{https://blog.neurips.cc/2025/09/30/reflections-on-the-2025-review-process-from-the-program-committee-chairs/}.

\bibitem[Charlin et~al.(2012)Charlin, Zemel, and Boutilier]{charlin_framework_2012}
Laurent Charlin, Richard~S. Zemel, and Craig Boutilier.
\newblock A {Framework} for {Optimizing} {Paper} {Matching}, February 2012.
\newblock URL \url{http://arxiv.org/abs/1202.3706}.
\newblock arXiv:1202.3706 [cs].

\bibitem[Cohan et~al.(2020)Cohan, Feldman, Beltagy, Downey, and Weld]{cohan_specter_2020}
Arman Cohan, Sergey Feldman, Iz~Beltagy, Doug Downey, and Daniel Weld.
\newblock {SPECTER}: {Document}-level {Representation} {Learning} using {Citation}-informed {Transformers}.
\newblock In Dan Jurafsky, Joyce Chai, Natalie Schluter, and Joel Tetreault, editors, \emph{Proceedings of the 58th {Annual} {Meeting} of the {Association} for {Computational} {Linguistics}}, pages 2270--2282, Online, July 2020. Association for Computational Linguistics.
\newblock \doi{10.18653/v1/2020.acl-main.207}.
\newblock URL \url{https://aclanthology.org/2020.acl-main.207/}.

\bibitem[Dumlao and Teplitskiy(2025)]{doi:10.1073/pnas.2507394122}
James M.~Zumel Dumlao and Misha Teplitskiy.
\newblock Geographical diversity of peer reviewers shapes author success.
\newblock \emph{Proceedings of the National Academy of Sciences}, 122\penalty0 (33):\penalty0 e2507394122, 2025.
\newblock \doi{10.1073/pnas.2507394122}.
\newblock URL \url{https://www.pnas.org/doi/abs/10.1073/pnas.2507394122}.

\bibitem[Ester et~al.(1996)Ester, Kriegel, Sander, and Xu]{ester_density-based_1996}
M.~Ester, H.~Kriegel, J.~Sander, and Xiaowei Xu.
\newblock In \emph{A {Density}-{Based} {Algorithm} for {Discovering} {Clusters} in {Large} {Spatial} {Databases} with {Noise}}, August 1996.
\newblock URL \url{https://dl.acm.org/doi/10.5555/3001460.3001507}.

\bibitem[Hanson et~al.(2023)Hanson, Barreiro, Crosetto, and Brockington]{Hanson2023TheSO}
Mark~A. Hanson, Pablo~G{\'o}mez Barreiro, Paolo Crosetto, and Dan Brockington.
\newblock The strain on scientific publishing.
\newblock \emph{Quantitative Science Studies}, 5:\penalty0 823--843, 2023.
\newblock URL \url{https://api.semanticscholar.org/CorpusID:263136473}.

\bibitem[Jerabkova et~al.(2023)Jerabkova, Patat, Primas, Dorigo, Sogni, Astolfi, Bierwirth, and Prümm]{jerabkova_first_2023}
Tereza Jerabkova, Ferdinando Patat, Francesca Primas, Dario Dorigo, Fabio Sogni, Lucas Astolfi, Thomas Bierwirth, and Michael Prümm.
\newblock The {First} {Results} of {Distributed} {Peer} {Review} at {ESO} {Show} {Promising} {Outcomes}, 2023.
\newblock URL \url{https://doi.eso.org/10.18727/0722-6691/5316}.
\newblock ISSN: 0722-6691 Publisher: European Southern Observatory (ESO).

\bibitem[Kerzendorf et~al.(2020)Kerzendorf, Patat, Bordelon, van~de Ven, and Pritchard]{kerzendorf_distributed_2020}
Wolfgang~E. Kerzendorf, Ferdinando Patat, Dominic Bordelon, Glenn van~de Ven, and Tyler~A. Pritchard.
\newblock Distributed peer review enhanced with natural language processing and machine learning.
\newblock \emph{Nature Astronomy}, 4\penalty0 (7):\penalty0 711--717, July 2020.
\newblock ISSN 2397-3366.
\newblock \doi{10.1038/s41550-020-1038-y}.
\newblock URL \url{https://www.nature.com/articles/s41550-020-1038-y}.
\newblock Publisher: Nature Publishing Group.

\bibitem[Khabsa et~al.(2015)Khabsa, Treeratpituk, and Giles]{khabsa_online_2015}
Madian Khabsa, Pucktada Treeratpituk, and C.~Lee Giles.
\newblock Online {Person} {Name} {Disambiguation} with {Constraints}.
\newblock \emph{Proceedings of the 15th ACM/IEEE-CS Joint Conference on Digital Libraries}, pages 37--46, June 2015.
\newblock \doi{10.1145/2756406.2756915}.
\newblock URL \url{https://dl.acm.org/doi/10.1145/2756406.2756915}.
\newblock Conference Name: JCDL '15: 15th ACM/IEEE-CS Joint Conference on Digital Libraries ISBN: 9781450335942 Place: Knoxville Tennessee USA Publisher: ACM.

\bibitem[Kovanis et~al.(2016)Kovanis, Porcher, Ravaud, and Trinquart]{Kovanis2016TheGB}
Michail Kovanis, Rapha{\"e}l Porcher, Philippe Ravaud, and Ludovic Trinquart.
\newblock The global burden of journal peer review in the biomedical literature: Strong imbalance in the collective enterprise.
\newblock \emph{PLoS ONE}, 11, 2016.
\newblock URL \url{https://api.semanticscholar.org/CorpusID:9484241}.

\bibitem[Langfeldt(2001)]{langfeldt2001decision}
Liv Langfeldt.
\newblock The decision-making constraints and processes of grant peer review, and their effects on the review outcome.
\newblock \emph{Social studies of science}, 31\penalty0 (6):\penalty0 820--841, 2001.

\bibitem[Latypova(2023)]{latypova2023decision}
Viktoriya Latypova.
\newblock Decision-making support in optimal multicriteria peer reviewer selection in scientific conference organization.
\newblock In \emph{2023 5th International Conference on Control Systems, Mathematical Modeling, Automation and Energy Efficiency (SUMMA)}, pages 383--386. IEEE, 2023.

\bibitem[Leyton-Brown et~al.(2022)Leyton-Brown, Mausam, Nandwani, Zarkoob, Cameron, Newman, and Raghu]{leyton-brown_matching_2022}
Kevin Leyton-Brown, Mausam, Yatin Nandwani, Hedayat Zarkoob, Chris Cameron, Neil Newman, and Dinesh Raghu.
\newblock Matching {Papers} and {Reviewers} at {Large} {Conferences}, August 2022.
\newblock URL \url{http://arxiv.org/abs/2202.12273}.
\newblock arXiv:2202.12273 [cs].

\bibitem[Merrifield and Saari(2009)]{merrifield_telescope_2009}
Michael~R. Merrifield and Donald~G. Saari.
\newblock Telescope {Time} {Without} {Tears}: {A} {Distributed} {Approach} to {Peer} {Review}.
\newblock \emph{Astronomy \& Geophysics}, 50\penalty0 (4):\penalty0 4.16--4.20, August 2009.
\newblock ISSN 13668781, 14684004.
\newblock \doi{10.1111/j.1468-4004.2009.50416.x}.
\newblock URL \url{http://arxiv.org/abs/0906.1943}.
\newblock arXiv:0906.1943 [astro-ph].

\bibitem[Mimno and McCallum(2007)]{mimno_expertise_2007}
David Mimno and Andrew McCallum.
\newblock Expertise modeling for matching papers with reviewers.
\newblock In \emph{Proceedings of the 13th {ACM} {SIGKDD} international conference on {Knowledge} discovery and data mining}, {KDD} '07, pages 500--509, New York, NY, USA, August 2007. Association for Computing Machinery.
\newblock ISBN 978-1-59593-609-7.
\newblock \doi{10.1145/1281192.1281247}.
\newblock URL \url{https://doi.org/10.1145/1281192.1281247}.

\bibitem[Nightingale and Scott(2007)]{nightingale2007peer}
Paul Nightingale and Alister Scott.
\newblock Peer review and the relevance gap: ten suggestions for policy-makers.
\newblock \emph{Science and Public Policy}, 34\penalty0 (8):\penalty0 543--553, 2007.

\bibitem[Olivo et~al.(2025)Olivo, Kerzendorf, Lu, Shields, Flörs, and Chen]{olivo2025practicalauthordisambiguationmetadata}
Vicente~Amado Olivo, Wolfgang Kerzendorf, Bangjing Lu, Joshua~V. Shields, Andreas Flörs, and Nutan Chen.
\newblock Practical author name disambiguation under metadata constraints: A contrastive learning approach for astronomy literature, 2025.
\newblock URL \url{https://arxiv.org/abs/2511.10722}.

\bibitem[Olivo et~al.(2026)Olivo, Jerabkova, Klencki, Carpenter, Malički, Patat, Strolger, and Kerzendorf]{olivo2026traditionalstatisticalrepresentationsoutperform}
Vicente~Amado Olivo, Tereza Jerabkova, Jakub Klencki, John Carpenter, Mario Malički, Ferdinando Patat, Louis-Gregory Strolger, and Wolfgang Kerzendorf.
\newblock Traditional statistical representations outperform generative ai in identifying expert peer reviewers, 2026.
\newblock URL \url{https://arxiv.org/abs/2605.18752}.

\bibitem[OpenReview(2025)]{noauthor_openreviewopenreview-expertise_2025}
OpenReview.
\newblock openreview/openreview-expertise, November 2025.
\newblock URL \url{https://github.com/openreview/openreview-expertise}.
\newblock original-date: 2018-08-02T18:12:33Z.

\bibitem[Pedregosa et~al.(2018)Pedregosa, Varoquaux, Gramfort, Michel, Thirion, Grisel, Blondel, Müller, Nothman, Louppe, Prettenhofer, Weiss, Dubourg, Vanderplas, Passos, Cournapeau, Brucher, Perrot, and Duchesnay]{pedregosa_scikit-learn_2018}
Fabian Pedregosa, Gaël Varoquaux, Alexandre Gramfort, Vincent Michel, Bertrand Thirion, Olivier Grisel, Mathieu Blondel, Andreas Müller, Joel Nothman, Gilles Louppe, Peter Prettenhofer, Ron Weiss, Vincent Dubourg, Jake Vanderplas, Alexandre Passos, David Cournapeau, Matthieu Brucher, Matthieu Perrot, and Edouard Duchesnay.
\newblock Scikit-learn: {Machine} {Learning} in {Python}, June 2018.
\newblock URL \url{http://arxiv.org/abs/1201.0490}.
\newblock arXiv:1201.0490 [cs].

\bibitem[Stelmakh et~al.(2019)Stelmakh, Shah, and Singh]{stelmakh_peerreview4all_2019}
Ivan Stelmakh, Nihar~B. Shah, and Aarti Singh.
\newblock {PeerReview4All}: {Fair} and {Accurate} {Reviewer} {Assignment} in {Peer} {Review}, November 2019.
\newblock URL \url{http://arxiv.org/abs/1806.06237}.
\newblock arXiv:1806.06237 [stat].

\bibitem[Stelmakh et~al.(2021)Stelmakh, Shah, Singh, and Daum{\'e}~III]{stelmakh2021novice}
Ivan Stelmakh, Nihar~B Shah, Aarti Singh, and Hal Daum{\'e}~III.
\newblock A novice-reviewer experiment to address scarcity of qualified reviewers in large conferences.
\newblock In \emph{Proceedings of the AAAI Conference on Artificial Intelligence}, volume~35, pages 4785--4793, 2021.

\bibitem[Stelmakh et~al.(2025)Stelmakh, Wieting, Xi, Neubig, and Shah]{stelmakh_gold_2025}
Ivan Stelmakh, John Wieting, Sarina Xi, Graham Neubig, and Nihar~B. Shah.
\newblock A {Gold} {Standard} {Dataset} for the {Reviewer} {Assignment} {Problem}, May 2025.
\newblock URL \url{http://arxiv.org/abs/2303.16750}.
\newblock arXiv:2303.16750 [cs].

\bibitem[Subramanian et~al.(2021)Subramanian, King, Downey, and Feldman]{subramanian_s2and_2021}
Shivashankar Subramanian, Daniel King, Doug Downey, and Sergey Feldman.
\newblock {S2AND}: {A} {Benchmark} and {Evaluation} {System} for {Author} {Name} {Disambiguation}.
\newblock \emph{2021 ACM/IEEE Joint Conference on Digital Libraries (JCDL)}, pages 170--179, September 2021.
\newblock \doi{10.1109/JCDL52503.2021.00029}.
\newblock URL \url{https://ieeexplore.ieee.org/document/9651767/}.
\newblock Conference Name: 2021 ACM/IEEE Joint Conference on Digital Libraries (JCDL) ISBN: 9781665417709 Place: Champaign, IL, USA Publisher: IEEE.

\end{thebibliography}

\end{document}